# Generalizing Morley's and Various Theorems with Realizability Computations

ERIC BRAUDE

Boston University

## Abstract

An approach is shown that proves various theorems of plane geometry in an algorithmic manner. The approach affords transparent proofs of a generalization of Morley's theorem and other well-known results by recasting them in terms of constraint satisfaction.

## 1  Introduction, Terminology, and Notation

This paper presents a perspective on figures in plane geometry by means of which various theorems can be systematically proved. The theorem on which this perspective is based, described in Section 2.3, has a particularly useful corollary concerning the matching of angles subtended by vertices. The approach shares some of the objectives of Wu's method ([10] and [11]) for automation in plane geometry proofs; but the latter translates plane figures into polynomial equations via Cartesian coordinates, which is different from the approach in the present work.

Morley's Theorem is an instance generalized by the approach we describe. It has interested many researchers, including Connes, Conway, Dijkstra, and Lebesgue. Connes [1] proved the theorem as "a group theoretic property of the action of the affine group on the line." Conway ([2] and [3]) called his proof "undisputedly simplest" (JCo2). Dijkstra gave a short "simple" proof in [4], a critique of this proof in [5], and a note on a tacit assumption in his proof [6].  Oakley and Baker [7] published an extensive bibliography in 1978. New proofs continue to appear (e.g., Stonebridge [9]) but the interest in Morley's theorem lies in the proof methods that it inspires.

The approach of the present work is based on real-valued mappings from the angles of plane figures. To this end, we standardize the figures under consideration: we define a *simple triangulated plane figure* (STPF) as a connected plane figure consisting of a finite set $\mathcal{S}$ of non-degenerate triangles such that for every vertex $v$ and triangle $T$ in $\mathcal{S}$, $v$ is either a vertex of $T$ or else external to $T$. This disallows vertices of one triangle occurring on the side of another except at one of the latter's vertices. We distinguish between an "angle"—a component of a triangle—and its size but when there is no chance of ambiguity, we sometimes conflate these.

Let $A_\mathscr{S}$ be the set of angles in the triangles of an STPF $\mathscr{S}$ (i.e., not simply the value of the angles). A mapping $m$ from $A_\mathscr{S}$ into the positive reals will be said to *realize* $\mathscr{S}$ if a plane figure exists for which the relationships among its triangles are the same as those for $\mathscr{S}$, and its angles have sizes equal to the corresponding values of $m()$. We will restrict our attention, without significant compromise, to convex STPF ("CSTPF's"). Figure 1 shows a non-realizing but otherwise well-behaved mapping on a CSTPF. (When discussing Morley's theorem, it has been common to express angle sizes in degrees, and this paper follows suit.)

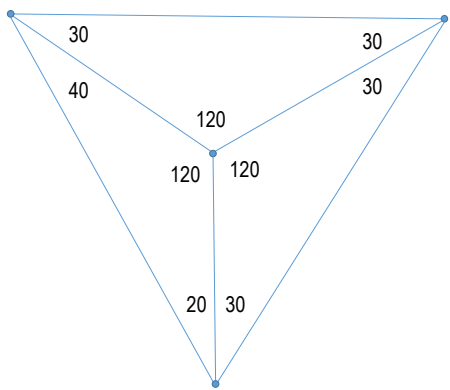

**Figure 1: Example of a non-realizing mapping on a CSTPF (convex simply triangulated plane figure)**

The vertices of a triangle $T$ will be denoted $T(1)$, $T(2)$, and $T(3)$. The set of triangles in an STPF $\mathscr{S}$ containing vertex $v$ will be denoted $\mathscr{T}_\mathscr{S}(v)$—or, when unambiguous, $\mathscr{T}(v)$. For vertices $v$ and $w$ of a triangle in a concrete instance of a STPF, $l(v, w)$ will denote the length of the corresponding line.

This paper uses triangulated forms for plane figures—CSTF's—as a standardized format with which to establish theorems and to systematically validate their proofs. A related problem is tessellation via Delaunay triangles but the motivation there is different (see, for example, [8]).

## 2  Realizability Theorem

The key theorem of this paper is as follows:

<u>Theorem 1 ("Realizability")</u>: A mapping from a convex simple triangulated figure $\mathscr{S}$ to the reals realizes $\mathscr{S}$ iff it satisfies the "π" condition, the "2π" condition, and the "alternating sine" condition, as defined in Section 2.1 below.

The idea for this theorem suggested itself to the author from Dijkstra's proof of Morley's Theorem [4], as well as a similar result on Delaunay triangulations [8]. The proof is established in Sections 2.1 and 2.2 below.

## *2.1*  **Necessary Conditions for Realizability**

Suppose that $m$ realizes CSTF $\mathcal{S}$.

1. ("$\underline{\pi \text{ condition}}$"): For every triangle $T$ in $\mathcal{S}$, $m(T(1)) + m(T(2)) + m((3)) = 180$ by elementary plain geometry.

2. ("$\underline{2\pi \text{ condition}}$"): For every vertex $v$ in $\mathcal{S}$, either $A(v) \leq 180$ or $A(v) = 360$, where $A(v)$ is defined as $\Sigma\{m(T(j)): T \in \mathcal{S} \text{ and } T(j) = v\}$.

Because $\mathcal{S}$ is convex, the $2\pi$ condition follows by considering vertices $v$ on the perimeter of $\mathcal{S}$ ($A(v) \leq 180$, $v$ "exterior") separately from those not ($A(v) = 360$, $v$ "interior").

3. ("$\underline{\text{Sine Rotation condition}}$"):

For every interior vertex $v$ with $\mathcal{A}(v) = \{T_1, T_2, \ldots, T_n\}$, $v = T_1(3) = T_2(3) = \ldots = T_n(3)$, $T_1(2) = T_2(1)$, $T_2(2) = T_3(1)$, $\ldots$, and $T_n(2) = T_1(1)$, we have:

$$\sin(m(T_1(1)))\cdot\sin(m(T_2(1)))\cdot \ldots \cdot\sin m(T_n(1))) = \sin(m(T_1(2)))\cdot\sin(m(T_2(2)))\cdot \ldots \cdot\sin(m(T_n(2)))$$

The setting for this condition is illustrated in Figure 2.

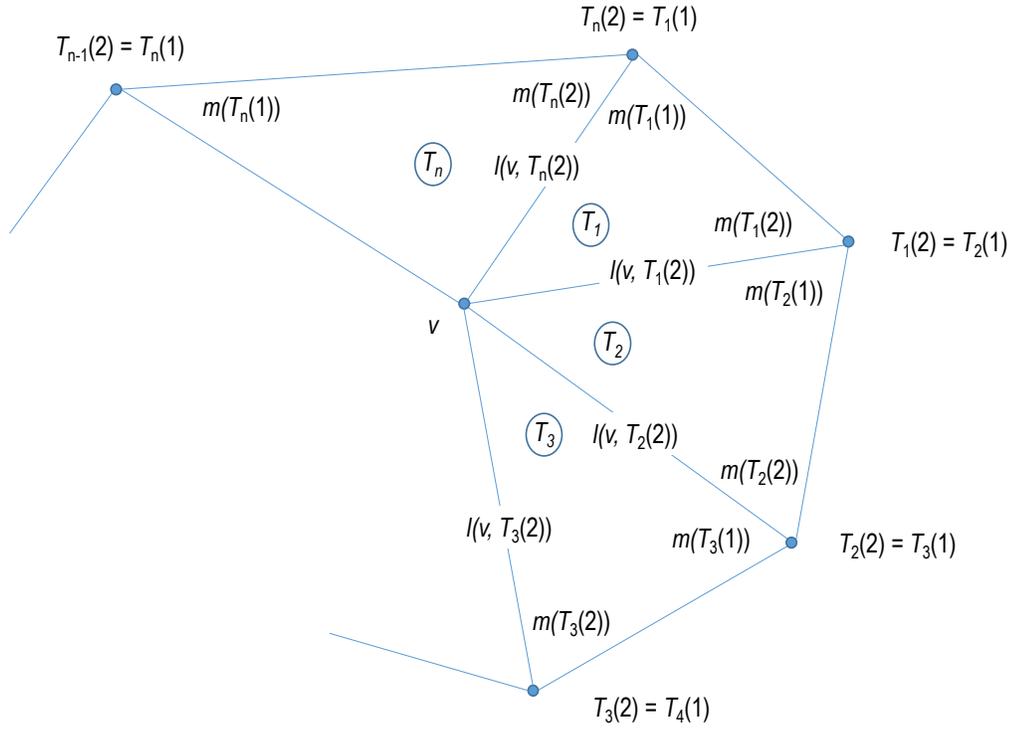

Using the sine rule,

$l(v, T_1(2)) / \sin(m(T_2(2))) = l(v, T_2(2)) / \sin(m(T_2(1)))$ and

$l(v, T_2(2)) / \sin(m(T_3(2))) = l(v, T_3(2)) / \sin(m(T_3(1)))$

… and

$l(v, T_n(2)) / \sin(m(T_1(2))) = l(v, T_1(2)) / \sin(m(T_1(1)))$.

From these we can conclude:

$l(v, T_1(2)) = l(v, T_2(2))) \sin(m(T_2(2))) / \sin(m(T_2(1))) =$

$l(v, T_3(2))) \sin(m(T_2(2))) \sin(m(T_3(2))) / \sin(m(T_2(1))) \sin(m(T_3(1))) = \ldots =$

$l(v, T_1(2))) \sin(m(T_2(2))) \sin(m(T_3(2))) \ldots \sin(m(T_n(2))) \sin(m(T_1(2)))$

$/ \sin(m(T_2(1))) \sin(m(T_3(1))) \ldots \sin(m(T_n(1))) \sin(m(T_1(1)))$

—and the sine rotation condition follows.

The sine rotation formula itself has been observed in various contexts (see, for example, [8] p110).

## *2.2* Sufficient Conditions for Realizability

To demonstrate the sufficiency of the $\pi$-, $2\pi$-, and sine rotation conditions, we provide a procedure for constructing a plane figure consistent with $\mathcal{S}$ and $m$. The procedure consists of realizing $\mathcal{A}(v)$ for every internal vertex $v$, and then doing the same for every external vertex. The procedure maintains the convexity of the figure realized at the completion of each of these steps.

For each internal vertex $v$, let $\mathcal{R}$ be the already-realized subset of $\mathcal{A}(v)$ and $T$ an unrealized triangle in $\mathcal{A}(v)$ that shares side $(v, r)$ with a triangle in $\mathcal{R}$. Let $x$ denote the required vertex of $T$ not on $(v, r)$. Because of convexity and the fact that $\Sigma\{m(T(j)): T \in \mathcal{R} \text{ and } T(j) = v\} < 360$, this can be done if $x$ is not already realized. Otherwise, we must show that the already-realized vertex $x$ is precisely the one required to realize $T$. (For example, in the mapping shown in the CSTF in Figure 3, it is simple to realize $T_1$ and $T_2$ but there is no freedom in realizing triangle $T_3$ because $v_3$ has already been realized.)

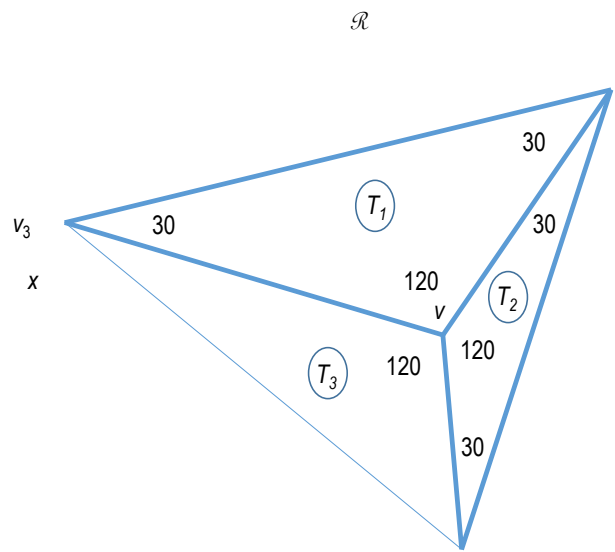

**Figure 3: Example of "last triangle" realization**

Figure 4 shows this "last triangle" problem in general, where triangle $T_{n-1}$ can be readily realized but $T_n$ can be constructed only by joining already-realized vertices $p$ and $q$.

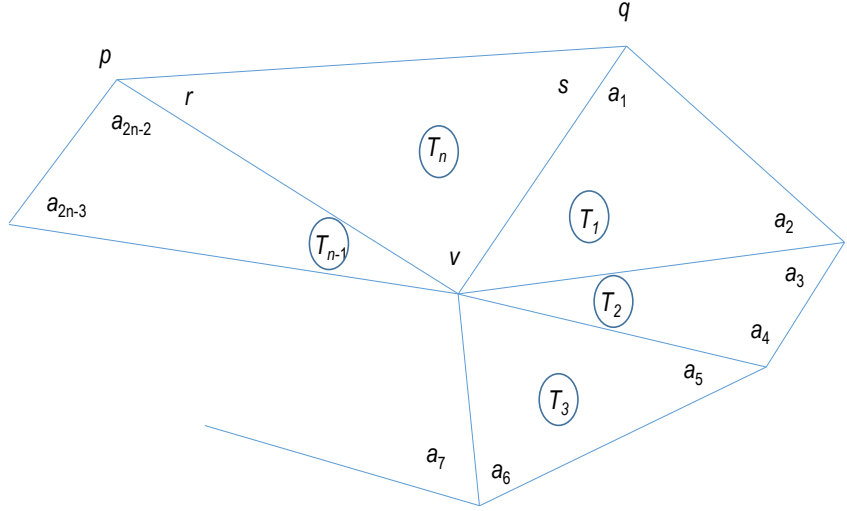

From Figure 4 and the argument Section 2.1, we can infer that

$$\sin(a_1)\,\sin(a_3)\,\ldots\,\sin(a_{2n-3})\,\sin(r) = \sin(a_2)\,\sin(a_4)\,\ldots\,\sin(a_{2n-2})\,\sin(s) \tag{1}$$

From the sine rotation condition, we have

$$\sin(a_1)\,\sin(a_3)\,\ldots\,\sin(a_{2n-3})\,\sin(a_{2n-1}) = \sin(a_2)\,\sin(a_4)\,\ldots\,\sin(a_{2n-2})\,\sin(a_{2n}) \tag{2}$$

Thus,

$$\sin(r)\,/\,\sin(s) = \sin(a_{2n-1})\,/\,\sin(a_{2n}). \tag{3}$$

We also know

$$r + s = a_{2n-1} + a_{2n} \tag{4}$$

Assuming that $a_{2n-1} >= r$, and defining $\phi$ as $a_{2n-1} - r$ (otherwise as $a_{2n} - s$), equation (3) becomes

$$\sin(r)\,\sin(s - \phi) = \sin(s)\,\sin(r + \phi) \tag{5}$$

Using a modification of a calculation used by Dijkstra and Ambuj Singh in [5], we obtain

$$\sin(r)[\sin(s)\,\cos(\phi) - \cos(s)\,\sin(\phi)] = \sin(s)[\sin(r)\,\cos(\phi) + \cos(r)\,\sin(\phi)]$$

thus

$$\sin(r)\cos(s)\sin(\phi) + \sin(s)\cos(r)\sin(\phi) = 0$$

$$\sin(r+s)\sin\phi = 0 \tag{6}$$

Since $0 < r + s < 180$, it follows that $\sin\phi = 0$. Since $0 <= \phi < 180$, we can conclude $\phi = 0$, $r = a_{2n-1}$, and $s = a_{2n}$. Thus, joining $p$ and $q$ does indeed realize $t_n$.

The realization obtained after performing this process on all internal vertices, is convex. Otherwise, a straight line would exist that does not intersect the triangles of $\mathcal{S}$ except at distinct external vertices $v_1$ and $v_2$. A sequence $v_1 = w_1, w_2, w_3, \ldots, w_{n-1}, w_n = v_2$ of external vertices would exist where $(w_1, w_2)$, $(w_2, w_3)$, $\ldots, (w_{n-1}, w_n)$ are sides in $\mathcal{S}$, and the $\pi$ condition would be violated by at least one element of $\{w_1, w_2, w_3, \ldots, w_{n-1}, w_n\}$.

It remains to realize the external vertices. For each such vertex $v$, we realize each unrealized triangle in $\mathcal{A}(v)$ by selecting one—$T$, say—which shares a side with a realized triangle. Suppose that this side has vertices $T(0) = v$, and $T(1)$. Because the $\pi$-condition applies to $T(0)$ and to $T(1)$, $T(2)$ must lie in the shaded region shown in Figure 5. The latter cannot impinge on any realized triangle, otherwise the figure realized so far would be concave. Thus, $T$ can be realized and so can all remaining triangles in $\mathcal{S}$.

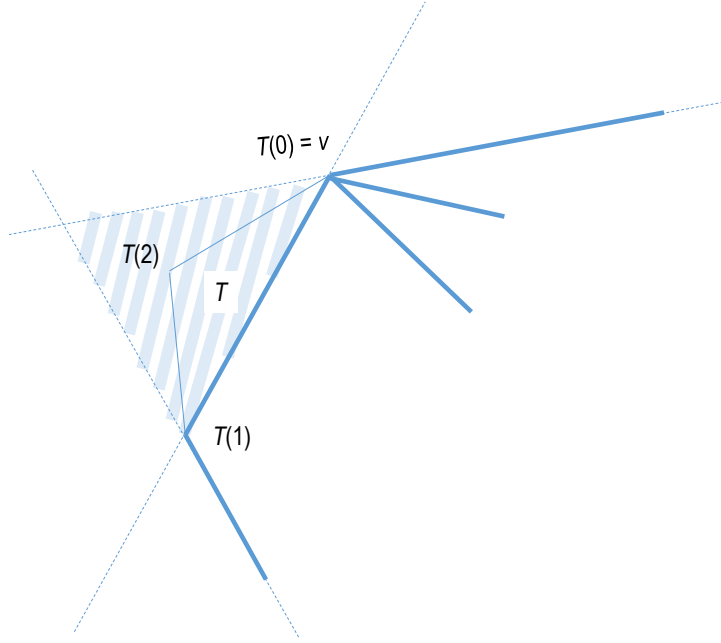

*T*(0) = *v*

*T*(2)

*T*

*T*(1)

The proof given at this point establishes that, given a mapping *m* on a CSTF satisfying the π-, 2π-, and sine rotation conditions, a figure can be constructed that is consistent with *m*. We are also assuming that *m* determines a unique plane figure (i.e., up to similarity). As proved in an essay on reasoning written by Dijkstra [6], it follows that the unique plane figure is precisely the constructed one. Dijkstra wrote [6] to explicate the reasoning for his proof of Morley's theorem. He did not point out the realizability theorem (Theorem 1), or the pairing corollary given below, however.

Hiroshima and Sugihara ([8] p 112) established theorems for Delauney triangulations similar to the realizability theorem but their motivation concerns tessellations rather than the subject of this paper.

## 2.3 Pairing Corollary

A simple, but very useful corollary follows immediately from the realizability theorem. For an internal vertex *v* in CSTF $\mathcal{S}$, suppose that $T_1$, $T_2$, ... , $T_n$, is the sequence of triangles in $\mathcal{S}$ containing *v*, where $T_i$ and $T_{i+1}$ share a side, and the angles of $T_i$ not at *v*, in clockwise order, are $a_{2i+1}$ and $a_{2i+2}$, as is shown in Figure 6. We define sets *odd*(*v*) and *even*(*v*) as $\{a_1, a_3, …, a_{2n-1} \}$ and $\{ a_2, a_4, …, a_{2n} \}$ respectively.

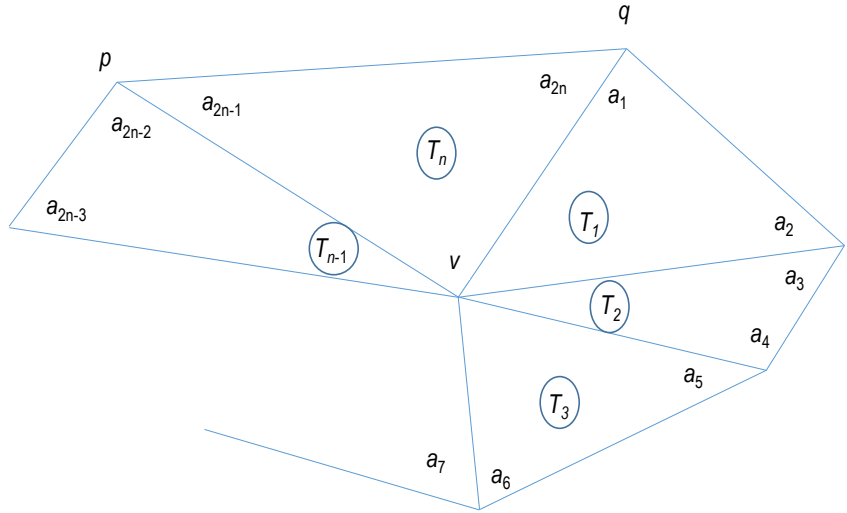

"Pairing" Corollary: A mapping from a CSTF to the reals is realizable if it satisfies the $\pi$ condition, the $2\pi$ condition, and if, for every internal vertex $v$, $odd(v)$ and $even(v)$ are identical multisets.

We will use the realizability theorem or, much more commonly, the pairing corollary in the following theorem-proving process, which we will refer to as a "realizability argument."

1. A set $P$ of properties of a CSTF $\mathcal{S}$ are given which uniquely determine $\mathcal{S}$ (i.e., up to similarity).
2. Using the realizability theorem or (more commonly) the pairing corollary, a realizable mapping $m$ from $\mathcal{S}$ to the reals is determined that satisfies $P$.
3. It is concluded that the angle sizes of $\mathcal{S}$ are precisely those specified by $m$.

Theorems can be generated by identifying property sets $P$ that uniquely determine $\mathcal{S}$. Conceivably, this identification process could be automated to an extent. Given a machine representation of a CSTF, the conditions of the pairing corollary can be machine-checked and thus the theorem "If $P$ then ..." proved if the check is successful. Several examples of realizability arguments are given below.

In figures, will use underlining to denote the assumed properties ("$P$" in the above).

# 3   A Generalization of Morley's Theorem

Besides being proved by the pairing corollary (below), Morley's Theorem is also suggested by it, as follows. If we assume that the latter were to be applicable, it would require {AUW, UVW} = {WUV, WVB} in Figure 7. Symmetry suggests that AUW does not pair with WUV. Thus, the pairings AUW = WVB and UVW = WUV are suggested, and the latter implies that the triangle is equilateral.

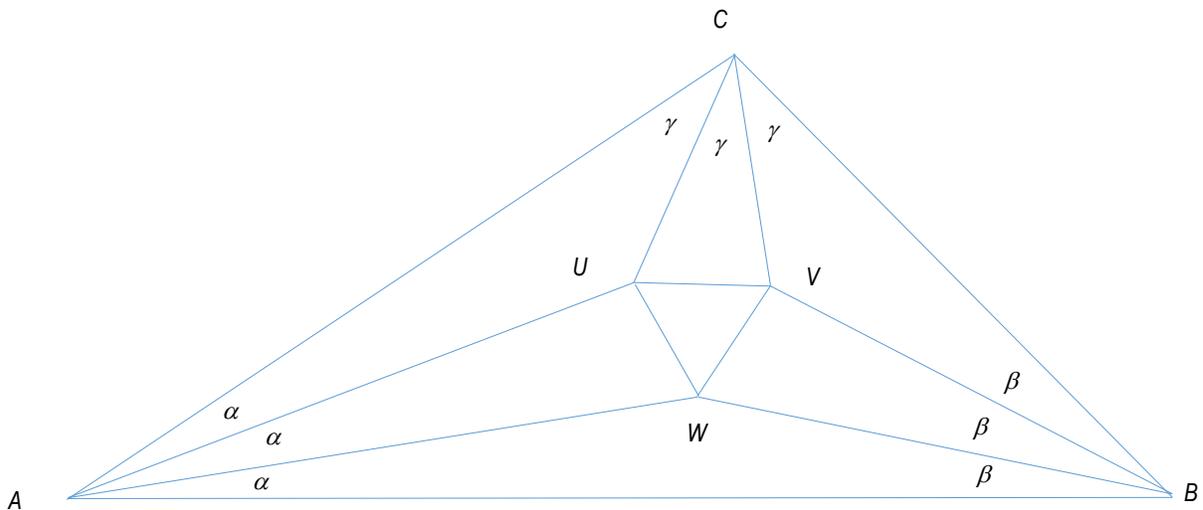

The proof technique described in this paper facilitates a generalization of Morley's Theorem. We define a *semi-regular* hexagon as a convex hexagon of the form $AD_1BD_2CD_3$, where $\angle D_1 = \angle D_2 = \angle D_3$. We will refer to $D_1$, $D_2$, and $D_3$ as the *constrained* vertices (or angles). There may be infinitely many realizations of a semi-regular hexagon thus specified.

Theorem 2 (Morley generalization): Let $A$, B, and $\Gamma$ be angles in (0, 180) with $A + B + \Gamma \geq 180$. There is a semi-regular hexagon with angles $A$, $\delta$, $B$, $\delta$, $\Gamma$, $\delta$ in which the trisectors of $A$, $B$, and $\Gamma$ form an equilateral triangle. Moreover, the latter's vertices lie on the bisectors of the constrained vertices.

Morley's theorem follows from this when $A + B + \Gamma = 180$ because in that case $\delta = 180$ and the hexagon is a triangle.

Note first that each of $A$, $B$, $\Gamma$, and $\delta$ does not exceed 180 and so the hexagon is convex. Let $A = 3\alpha$, $B = 3\beta$, and $\Gamma = 3\gamma$. Since $A + B + \Gamma + 3\delta = 720$, $\alpha + \beta + \gamma + \delta = 240$. Figure 8 shows a realization which conforms to the conditions of the pairing corollary, and hence proves Theorem 2.

**Figure 8: Proof of generalization of Morley's Theorem**

# 4  Theorem Families

In this section, we show how the realizability argument can be used to prove generalizations of well-known theorems, together with new ones. The approach actually provides (in fact, is required to provide) solutions for all angles.

## 4.1  Categorization of Triangle Concurrency

The pairing corollary suggests a categorization of various plane figures: CSTF's whose internal vertices conform to the corollary's premises. The simplest nontrivial case is a CSTF consisting of three triangles with a common vertex, as shown in Figure 9.

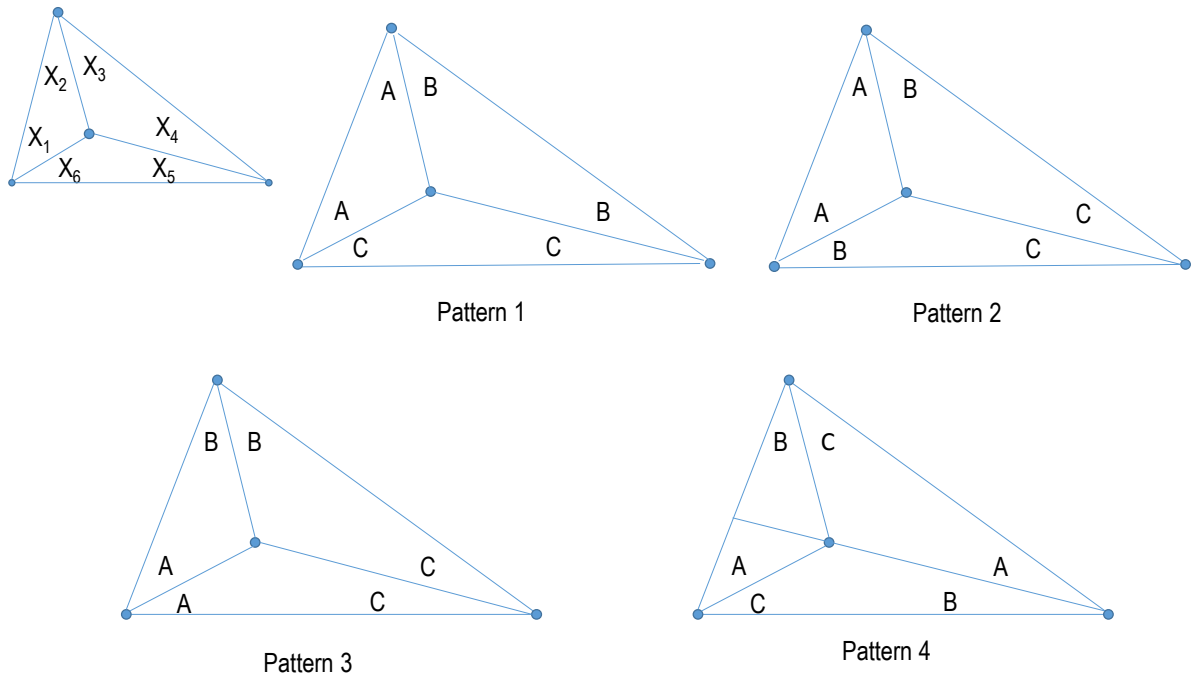

**Figure 9: Simplest nontrivial CSTF**

The four patterns for pairing among angles $X_1$ through $X_6$ are listed in Table 1. The primary organization is a binary one ($s$ = "same angles in the triangle"; $d$ = "different …").

**Table 1: Patterns for internal vertices with 3 triangles**

| *s s s* | *s d d* | *d d d* |
|---|---|---|
| 1. AA BB CC | 2. XX YZ ZY | 3. AB BC CA |
| | | 4. AB CA BC |

Pattern 1 describes the triangle's incenter. Pattern 2 expresses a concurrency from an angle bisector in an isosceles triangle. Pattern 3 expresses and proves the concurrency of angle bisectors. In Pattern 4, the concurrent lines are perpendiculars since $A + B + C = 90$. Concurrency patterns for four triangles is discussed in Section 4.1.

## 4.2 Bisector Concurrency

Pattern 3 in Figure 9 yields the classical elementary bisector concurrency theorem. We return to semi-regular hexagons to generalize this.

Theorem 3: In every semi-regular hexagon, the bisectors are concurrent.

To see this, let *I* be the intersection of the bisectors at $\alpha$ and $\beta$. We claim that the (unique) CSTF obtained by joining *I* to the remaining vertices is realized in Figure 10, where $x^*$ denotes $180 - x/2 - \delta/2$. From the given conditions, $\alpha + \beta + \gamma + 3\delta = 4 \cdot 180$. The $\pi$- and alternating sine conditions are evident. To see that the $2\pi$ condition is valid at *I*, note that $\alpha^* + \beta^* + \gamma^* = 3 \cdot 180 - (\alpha + \beta + \gamma)/2 - 3\delta/2 = 3 \cdot 180 - 2 \cdot 180 = 180$.

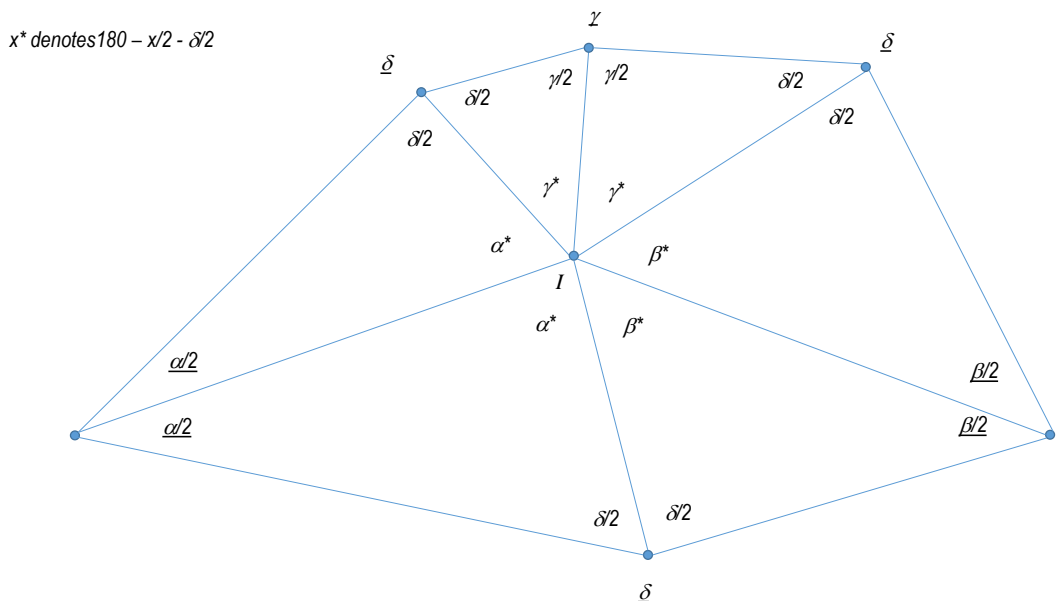

**Figure 10: Proof of Theorem 4**

## 4.3 Median Concurrency

In this section we use the realizability theorem (i.e., not the pairing corollary) to generalize the concurrency theorem for a triangle's medians. This is done, for a triangle with angles $\alpha$, $\beta$, and $\gamma$, by considering line segments of the form $BB_0$ as shown in Figure 11, where $0 < \delta < \min(\alpha/2, \beta/2, \gamma/2)$. Segment $BB_0$, which we will refer to as a *semi*-median, and which is determined by $\beta_1$ and $\beta_2$, becomes a conventional median as $\delta$ tends to zero.

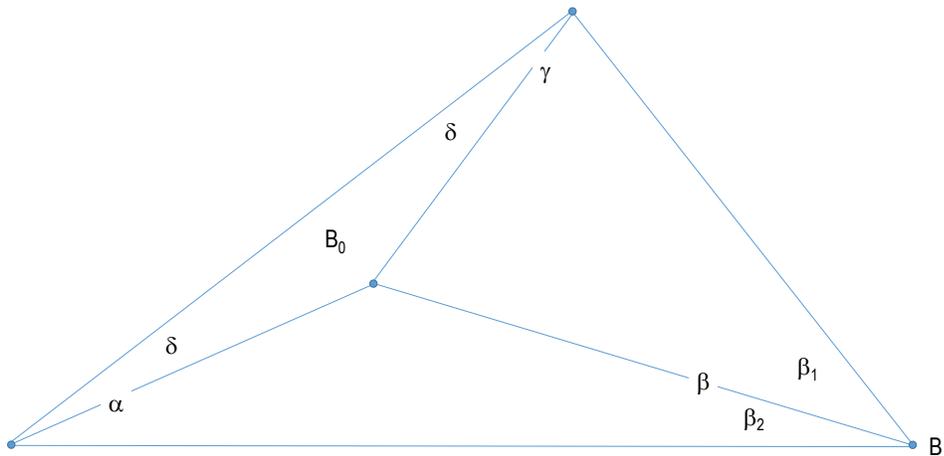

**Figure 11: Semi-medians**

From the realizability theorem (Theorem 1), $\sin(\beta_1)\sin(\alpha - \delta)\sin(\delta) = \sin(\beta_2)\sin(\delta)\sin(\gamma - \delta)$ and so $\sin(\beta_1) = \sin(\beta_2)\sin(\gamma - \delta)/\sin(\alpha - \delta)$. Similar equations hold for $\alpha$ and $\gamma$. Note that:

$\sin(\alpha_1)\sin(\beta_1)\sin(\gamma_1) =$

$\sin(\alpha_2)\sin(\beta - \delta)/\sin(\gamma - \delta) \cdot \sin(\beta_2)\sin(\gamma - \delta)/\sin(\alpha - \delta) \cdot \sin(\gamma_2)\sin(\beta - \delta)/\sin(\alpha - \delta)$

$= \sin(\alpha_2)\sin(\beta_2)\sin(\gamma_2)$.

Hence, the mapping shown in Figure 12 realizes the solid-line CSTF shown within. But this extends as shown in the figure as a whole, and the concurrency is proved.

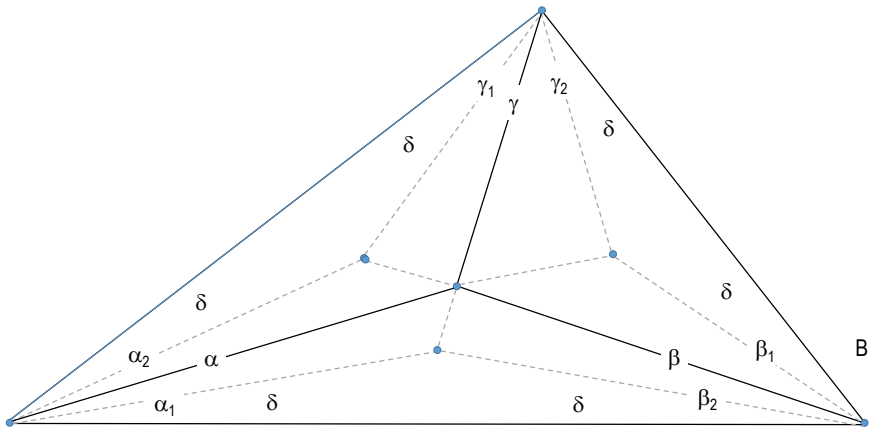

**Figure 12: Realization for Semi-medians**

The above proof of a generalization of the median–concurrency theorem can be contrasted with the traditional proof, which uses Ceva's theorem.

## 4.4 Morley Theorems

The realizability theorem, and especially its pairing corollary, can be used to generate new theorems by starting with CSTF's of interest, imposing realizations on them, and then selecting conditions sufficient to characterize these. For example, we can use the Morley CSTF to determine a point $V$ by means of four trisectors and an equilateral triangle, and conclude, using Figure 13 and a realizability argument, that $V$ is the intersection of the remaining two trisectors.

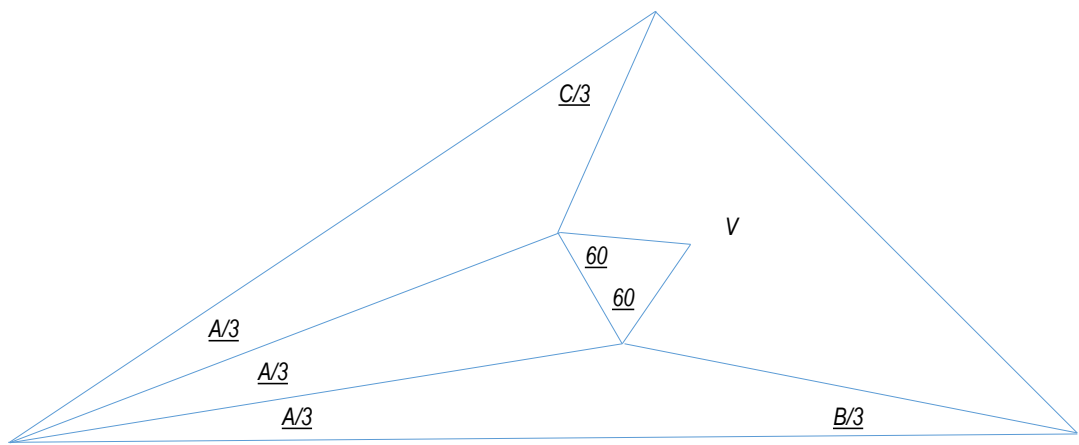

**Figure 13: A Morley-type theorem**

The following is a non-Morley equilateral theorem

Theorem 4: Let ABC be a triangle with incenter $I$, point $P$ on $AC$ with $AIP$ = B/2 + 60 and $Q$ on $BC$ with $BIQ$ = A/2 + 60. Then $IPQ$ is equilateral.

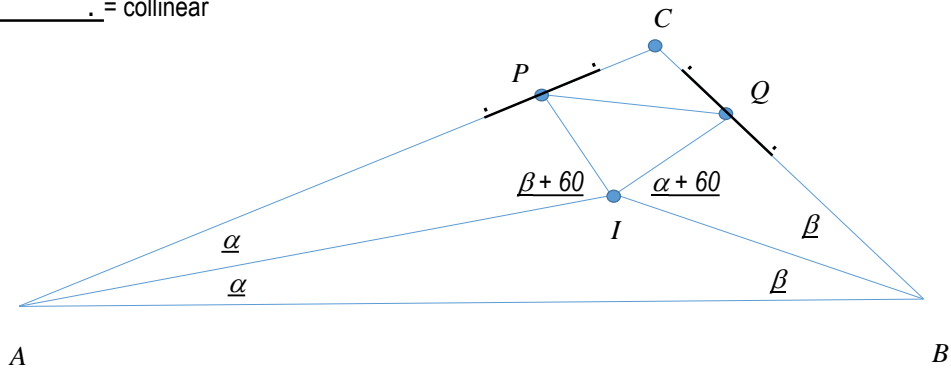

**Figure 14: Premises of Theorem 4**

The proof follows from Figure 15, which satisfies the conditions of the pairing corollary (in particular, at point *I*).

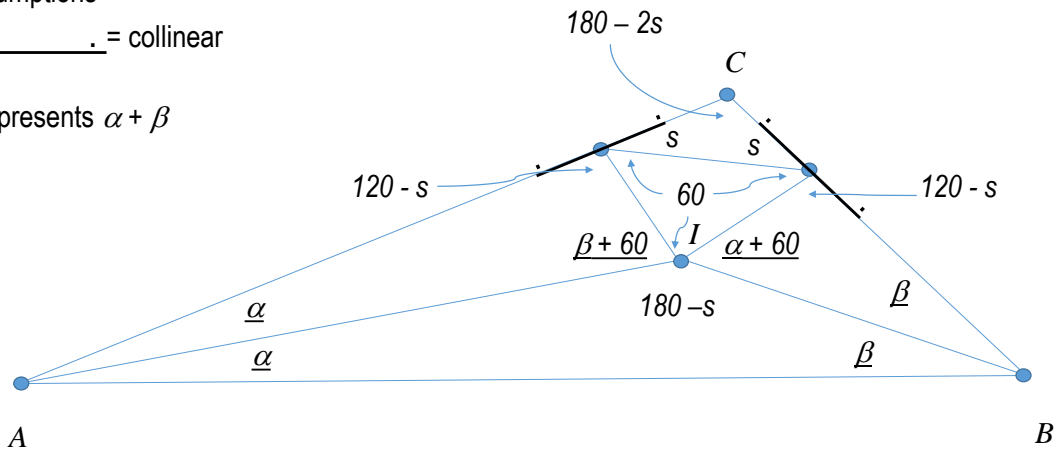

**Figure 15: Proof of Theorem 4**

## 4.5  Pappus' Theorems

The parallel versions of Pappus' theorem (also a special case of Pascal's Theorem) can be obtained from the CSTF shown in Figure 16, where parallelism is specified by the equality of corresponding angles, and α, β, and γ are independent.

**Figure 16: CSTF for Pappus' Parallel Theorems**

The proof of Pappus' parallel theorem—indeed, the entire angle solution to this figure—is shown in Figure 17 since it satisfies the conditions of the pairing corollary. In particular, it establishes the parallelism of AD and CF since their included angles are complementary.

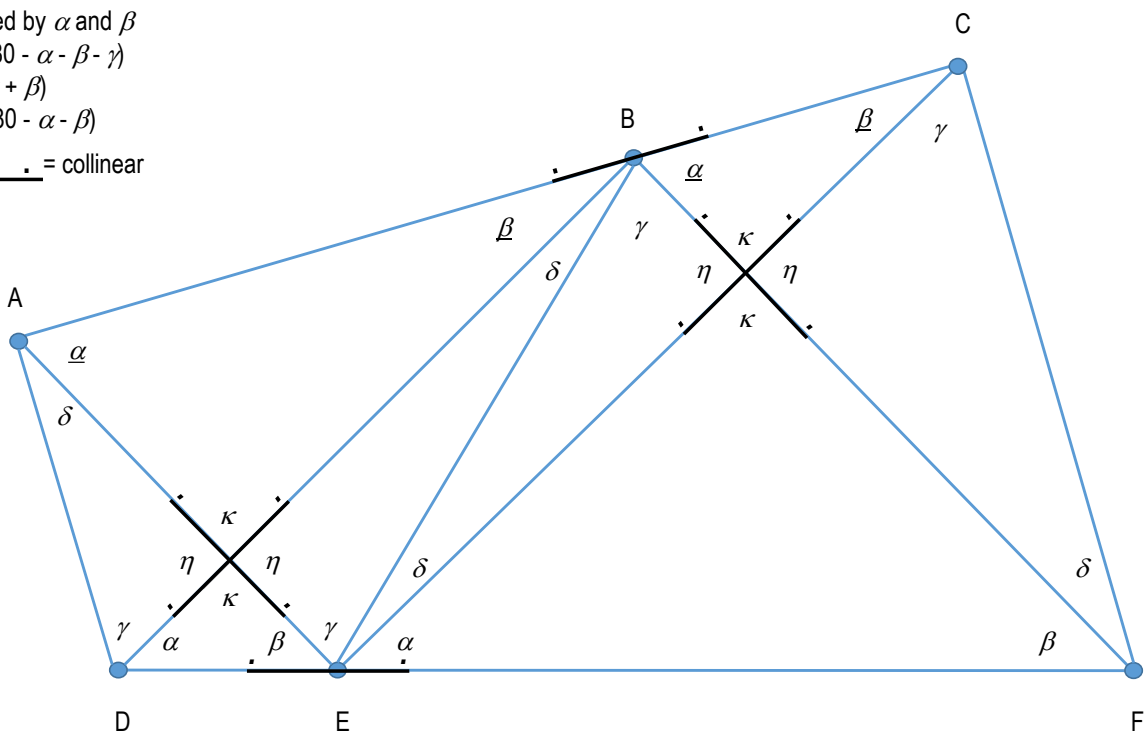

$\gamma$ is determined by $\alpha$ and $\beta$
$\delta$ denotes $(180 - \alpha - \beta - \gamma)$
$\eta$ denotes $(\alpha + \beta)$
$\kappa$ denotes $(180 - \alpha - \beta)$

•———————• = collinear

**Figure 17: Solution to Pappus' parallelism theorem**

Any set of conditions sufficient to determine the CSTF in Figure 17 creates a theorem whose conclusions are expressed by it. Figure 18 is an example. For example, we can conclude the following:

Theorem 5: If two corresponding vertices of two similar triangles lie on each other's corresponding sides, then the lines joining the corresponding vertices of these two sides (shown dotted in Figure 18) are parallel.

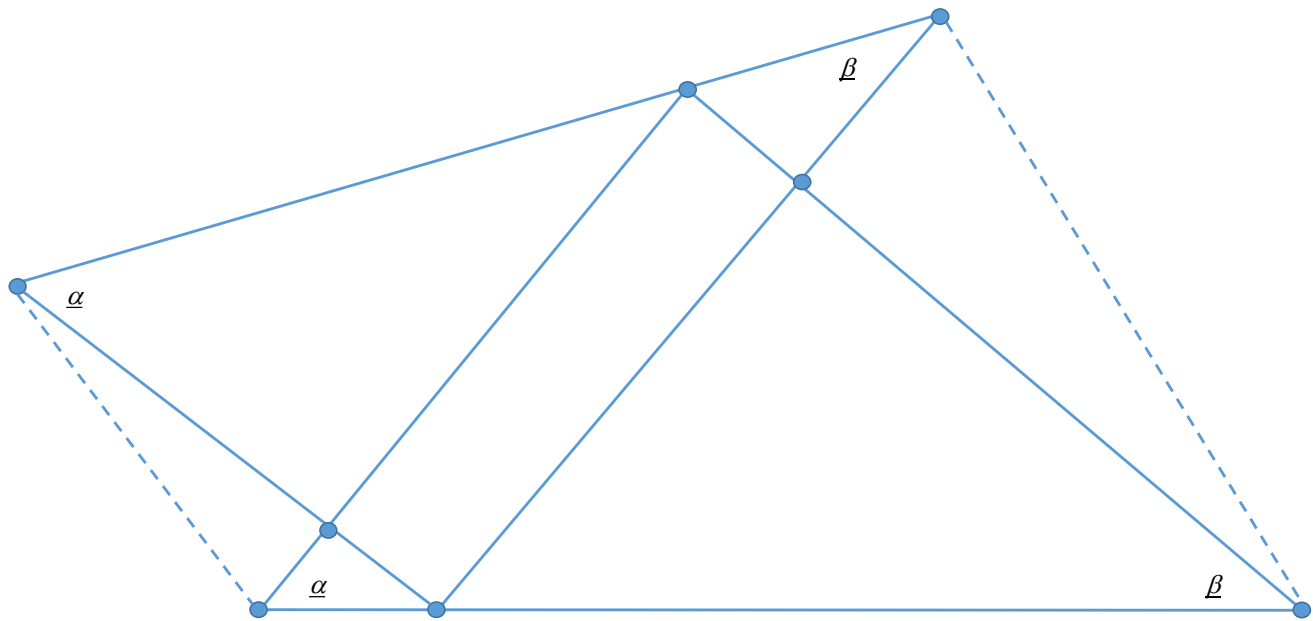

**Figure 18: Premises and conclusion of Theorem 6, a Pappus-like variation**

## 4.6 Other Theorems

In this section, we again use the pairing corollary to generate new theorems by starting with interesting CSTF's. The CSTF can be one that promises to continue a line of reasoning, such as the "quadriceptors" of triangles, as in Figure 19 below, or else a combination of familiar shapes, as in the second example below.

Continuing a line of reasoning from Morley's Theorem, suppose that that we divide each angle of a triangle into four equal parts, as in the underlined parts of Figure 19 (so that $\alpha + \beta + \gamma = 45$). The mapping shown satisfies the conditions of the pairing corollary and since the figure is a uniquely determined CSTF (up to similarity), they reflect its true angles. The repeated-asterisk notation is due to Conway [2] and, again, the proof can be mechanically verified with the pairing corollary.

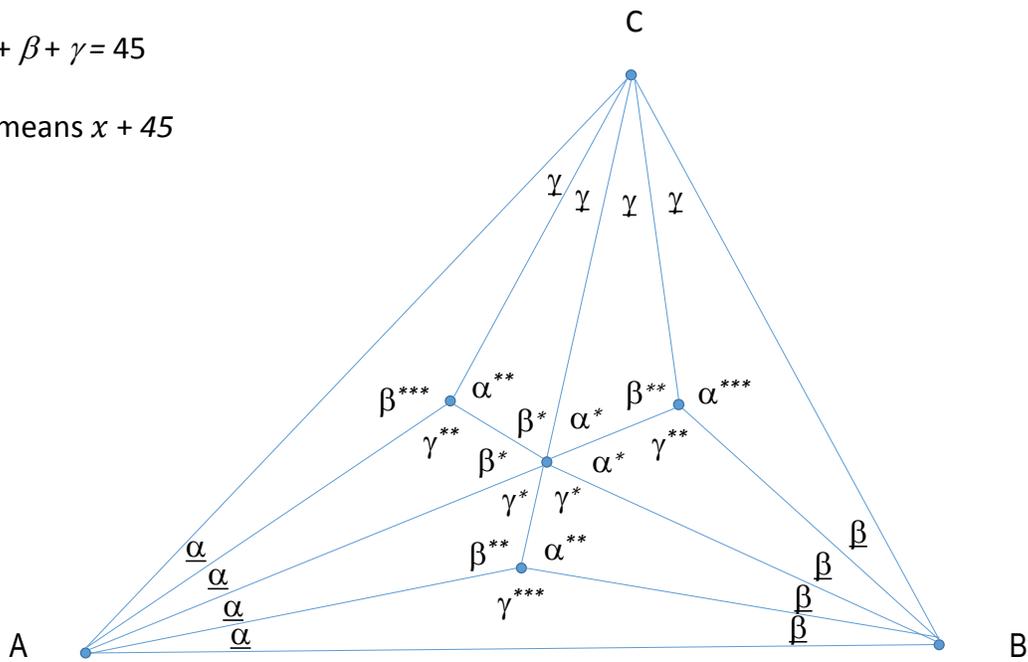

**Figure 19: Quadriceptors of a triangle**

As a final example, this one known in the literature, consider the CSTF and realization shown in Figure 20, where $I$ is the circle's center.

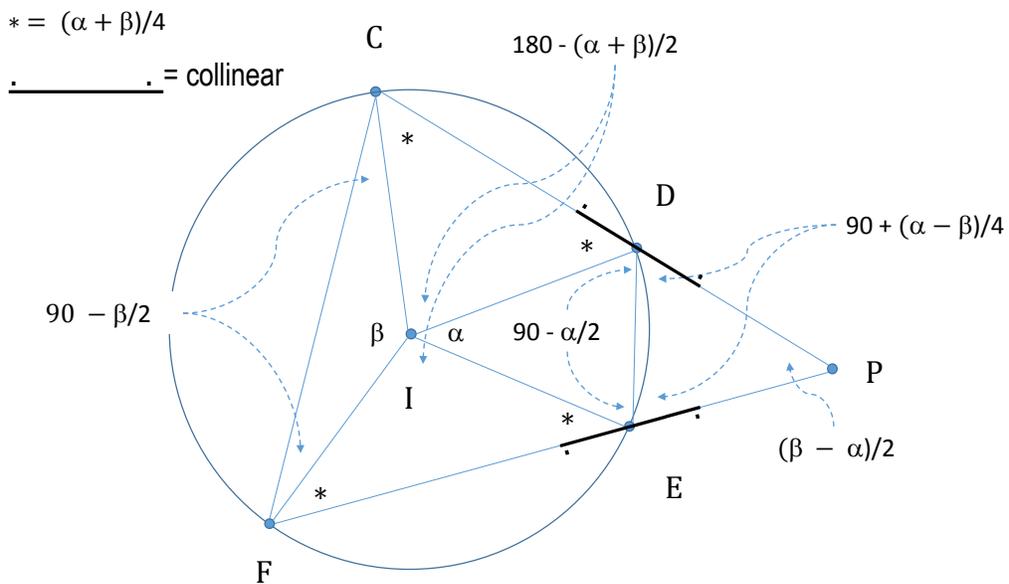

Theorems, with proofs, can be created from this realized CSTF, such as the following:

Theorem 6: Suppose that $C$, $D$, $E$, and $F$ are points on a circle where $EF$ subtends $\alpha$ at the center, $CD$ subtends $\beta$, $\alpha < \beta$, and $EC$ and $FD$ meet at $P$. Then $CPF = (\beta - \alpha)/2$.

# 5  Future Work and Conclusion

Besides the potential for machine check-ability of proposed proofs using this paper's approach, a systematic categorization of various plane figures suggests itself, discussed below.

## 5.1  Categorizing Plane Figures

We enquire whether a systematic categorization is possible of the CSTF's which satisfy the conditions of the pairing corollary. The categorization of the relevant three-triangle CSTF's shown in Figure 9 is a beginning. Continuing this for interior vertices of four triangles, we seek, in effect, to categorize the equivalence subsets of the set of all four pairs of letters from {A, B, C, D} in which each letter occurs exactly once as a first element and once as a second element, and the equivalence relationship is a bijection and/or string reversal. There are 10 such categories, as explained next.

Using $s$ ("same") to denote the fact that the relevant pair of angles are of the same in magnitude, and $d$ otherwise, the categorization can be divided into: an $ssss$-type element (i.e., of the form AA BB CC DD), an $sdsd$-type element, the two $sddd$-type elements (AA BC CD DB and AA BC DB CD—essentially the "different" options in Figure 9), and the six $dddd$-type elements illustrated in Figure 21.

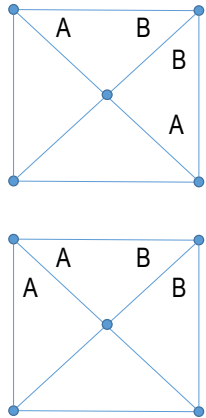
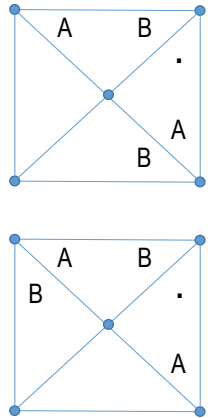
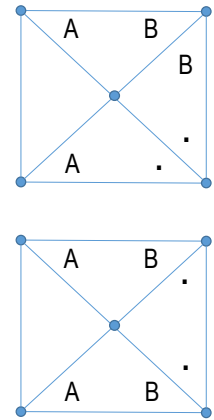

**Figure 21: Representative figures for *dddd* configurations**

Evidently, the categorization described can be mechanically generated for more complex figures.

## 5.2 Conclusion

A computational means for proving various theorems of plane geometry is shown. It affords a transparent proof of a generalization of Morley's and other well-known theorems by casting various plane geometry problems as constrained mappings. It facilitates the establishment of families of theorems in plane geometry consisting of complete angle solutions.